\documentclass[twocolumn,a4paper]{nveart}
\usepackage{epsfig,amssymb,pifont}

\begin{document}

\company{Utrecht University}
\journal{}

\begin{frontmatter}
\title{On the Use of Transverse Momentum Spectra to probe Thermalisation\\
       and Collective Expansion in Relativistic Nucleus-Nucleus Collisions}

\author{Nick van Eijndhoven}
\address{Department of Subatomic Physics, Utrecht University/NIKHEF\\
         P.O. Box 80.000, NL-3508 TA Utrecht, The Netherlands}

\begin{abstract}
Transverse momentum spectra of the reaction products from proton-proton and nucleus-nucleus
collisions are discussed within the phenomenological frameworks of statistical phase-space
and string fragmentation models. 
It will be shown that a string fragmentation model can accommodate all of the observations
discussed here and consequently it is claimed that no conclusion about the thermal nature
or collective expansion of the created system may be derived from observed transverse momentum spectra.
\end{abstract}

\begin{keyword}
Nucleus-nucleus collisions,
thermalisation,
transverse momentum spectra,
string fragmentation.
\PACS{13.75C, 13.85, 25.70.G, 24.10.P, 25.75.D}
\end{keyword}
\end{frontmatter}

\section{Introduction}
With the highly-relativistic heavy-ion physics programmes reaching maturity,
one can now start to address the question of
what are the observables to provide unambiguous evidence for the formation of
a Quark-Gluon Plasma (QGP) state.\\
Before answering this question, physicists have first tried to see
whether the current data indicate the formation of a thermalised system,
the logical implication of QGP formation.
In doing so, transverse momenta of the particles produced in the collisions
have been investigated.

In very energetic heavy-ion collisions a large number of particles is produced
in a relatively small volume and consequently many secondary collisions between the
reaction products take place (rescattering).
This makes experiments with highly relativistic heavy-ion beams well suited to search
for the formation of a thermally equilibrated system.

Several attemps \cite{becattini,pbm} have been made to investigate whether the spectra
and (relative) yields of the produced particles indicate the formation of a thermally
and/or chemically equilibrated system.\\
In addition, interpretation of the observed spectra within the framework of hydrodynamical
evolution of the created system followed by chemical and thermal freeze-out of the hadron source
has led to the idea of collective expansion effects reflected by the produced particles
\cite{heinz,na49qm97}.

In this paper it will be shown that particle production through the mechanism of string
fragmentation also leads to particle spectra compatible with the observations discussed here,
without the necessity of collective effects.  

\section{The thermal model}
A theoretical description of the momentum spectra of particles emerging from a thermalised system
is obtained by assuming that all the constituents obey a Boltzmann distribution.\\
In the case of formation of such a thermalised system by energetic collisions of heavy ions,
it is convenient to introduce the perpendicular and parallel momentum components with respect
to the direction of the incoming projectile.
These observables are indicated by $p_{\perp}$ and $p_{\parallel}$, respectively, whereas we
also introduce the transverse mass $m_{\perp} \equiv \sqrt{p_{\perp}^{2}+m^{2}}$.
The distribution of the emerging particles as a function of these variables is given by
\begin{equation}
 {\rm d}N(\vec{p},m,T) = A\,e^{-\sqrt{m_{\perp}^{2}+p_{\parallel}^{2}}/T}m_{\perp}
                         \,{\rm d}m_{\perp}{\rm d}p_{\parallel}{\rm d}\varphi ~,
\label{eq:boltzmann}
\end{equation}
where $T$ represents the temperature of the system and $A$ contains the normalisation to the total
number $N$ of particles as well as the (mass) terms which account for the proper dimensions.

\begin{figure}[htb]
\begin{center}
\resizebox{8.4cm}{!}{\includegraphics{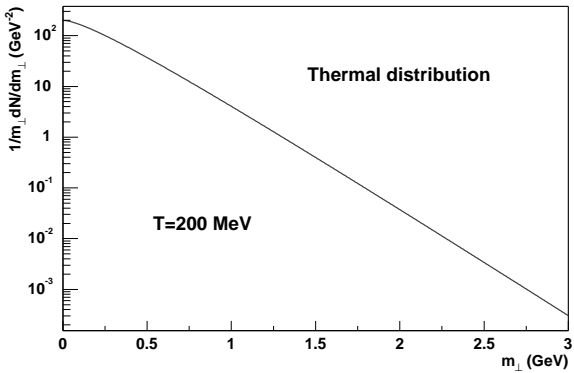}}
\end{center}
\caption{Calculated particle yield for a thermal source of 200~MeV temperature
         according to Eq.~(\ref{eq:thermal1}).}
\label{fig:thermal1}
\end{figure}

Using the azimuthal symmetry of the situation described above,
integration over $p_{\parallel}$ and $\varphi$
results in the following expression for the differential particle yields
\begin{equation}
 \frac{{\rm d}N(m_{\perp},T)}{m_{\perp}{\rm d}m_{\perp}} \sim m_{\perp}\,
 K_{1}\left( \frac{m_{\perp}}{T} \right) ~,
\label{eq:thermal1}
\end{equation}
where $K_{1}$ represents the modified Bessel function.\\
Making use of the relations $\displaystyle K_{1}(x \gg 1) \sim \sqrt{\frac{\pi}{2x}}\,e^{-x}$
and $K_{1}(x \ll 1) \sim x^{-1}$ we see from Eq.~(\ref{eq:thermal1}) that
at high momenta an essentially exponential behaviour of the particle distribution is
obtained, whereas at low momenta the differential yield saturates at a level proportional
to the temperature of the system.\\
The theoretical particle spectrum for a thermalised system at a temperature of 200~MeV
is shown in Fig.~\ref{fig:thermal1}.

According to the theoretical description of the particle spectra produced by a
thermalised system, the following observations have been made~: 
\begin{itemize}
\item Various particle species will show the same shape in the transverse mass
      spectra, the so-called $m_{\perp}$ scaling.
\item At high $p_{\perp}$ initial hard scattering of the constituents starts to play a role \cite{Cronin}
      and as such an enhancement at high $p_{\perp}$ is expected, the so-called Cronin effect. 
\item At low $p_{\perp}$ one starts to become sensitive to the products of resonance decays,
      which induces a rapidity--dependent enhancement \cite{resonances}.
\end{itemize}
From this it is seen that, in comparison with experimental results, the description of the
particle spectra as outlined above is only valid in a limited $m_{\perp}$ interval.

Furthermore, in general Eq.~(\ref{eq:thermal1}) can not be used in comparison with experimental
data because of the fact that, due to the limited phase-space coverage of the experimental
apparatus, only a limited range in $p_{\parallel}$ can be addressed.\\
To account for this effect, it is more convenient to express the particle yields in terms
of the particle rapidity
\begin{equation}
y=0.5 \ln \left( \frac{E+p_{\parallel}}{E-p_{\parallel}} \right) ~,
\label{eq:rapidity}
\end{equation}
where $E$ represents the particle energy.\\
Using the identity $p_{\parallel}=m_{\perp} \sinh(y)$ the Boltzmann distribution
of Eq.~(\ref{eq:boltzmann}) can be written as~:
\begin{equation}
 {\rm d}N(\vec{p},m,T) \sim
 e^{-m_{\perp}\cosh(y)/T}m_{\perp}^{2}\cosh(y)\,{\rm d}m_{\perp}{\rm d}y{\rm d}\varphi .
\label{eq:boltzmann2}
\end{equation}
Here it should be noted that $y$ represents the rapidity in the restframe of the thermalised
system, since all the descriptions so far have been performed in this frame.

Unfortunately, the particle distribution of Eq.~(\ref{eq:boltzmann2}) can not be evaluated
analytically to provide differential particle yields as a function of $m_{\perp}$ and $T$
within a certain rapidity range.
For each experimental situation numerical integration over the corresponding rapidity
coverage has to be performed in order to compare the model with the experimental data.

\section{What do the data tell us ?}
In this section the theoretical model described before will be tested against
experimental data.
For this purpose a selection of data obtained from $^{208}$Pb+$^{208}$Pb collisions at the CERN-SPS
with a beam energy of 158$A$~GeV and p+p collisions at the CERN-ISR at $\sqrt{s}$ of 23~GeV
has been used.\\
Here it should be realised that the $\sqrt{s}$ per nucleon-nucleon collision in case of the
CERN-SPS heavy-ion collisions is comparable to that of the CERN-ISR proton-proton collisions.

\begin{figure}[htb]
\begin{center}
\resizebox{8.5cm}{!}{\includegraphics{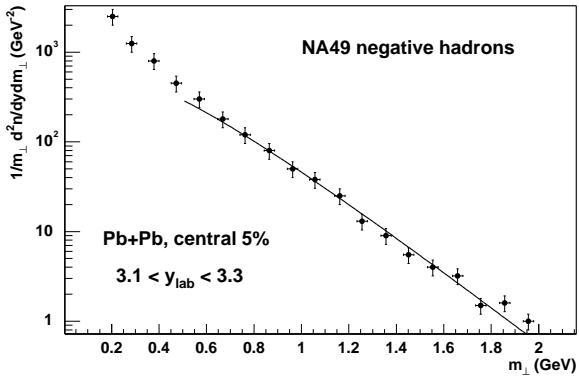}}
\end{center}
\caption{Charged hadron yields \cite{na49} as observed within the NA49 experiment.
         The solid line indicates a fit with the thermal model from Eq.~(\ref{eq:boltzmann2}).}
\label{fig:na49}
\end{figure}

\begin{figure}[htb]
\begin{center}
\resizebox{8.5cm}{!}{\includegraphics{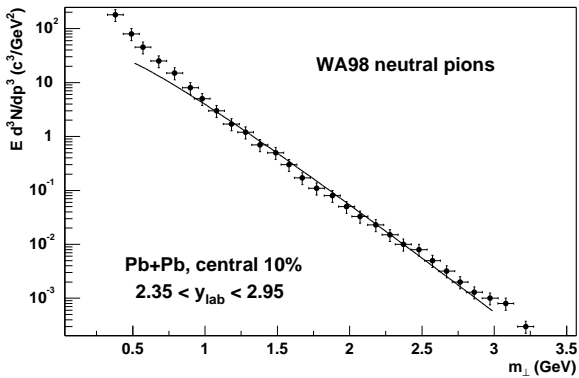}}
\end{center}
\caption{WA98 observed $\pi^{0}$ transverse mass spectrum \cite{wa98}.
         The solid line indicates a fit with the thermal model from Eq.~(\ref{eq:boltzmann2}).}
\label{fig:wa98}
\end{figure}

Concerning the CERN-SPS heavy-ion data we will investigate the charged hadron yields as
observed in the NA49 experiment \cite{na49} and the $\pi^{0}$ yields as observed
in the WA98 experiment \cite{wa98}. For both experiments the mid-rapidity value in the
laboratory system was 2.9.\\
The various spectra are shown in Figs.~\ref{fig:na49} and \ref{fig:wa98}, respectively.
The fact that the spectrum of Fig.~\ref{fig:wa98} can also be compared directly with
Eq.~(\ref{eq:boltzmann2}) is easily demonstrated by using the identities $E=m_{\perp}\cosh(y)$
and ${\rm d}p_{\parallel}=E{\rm d}y$.

Comparison of the observed spectra with our thermal model calculations reflected by Eq.~(\ref{eq:boltzmann2})
leads to the preliminary conclusion that indeed the spectra exhibit a thermal character compatible
with a temperature of $205$~MeV.
Furthermore, it has been observed \cite{na49qm97,na49} that the spectra of various particle species
exhibit larger inverse slopes with increasing particle mass. This phenomenon has been interpreted
in terms of a collective expansion of the hadronic source \cite{heinz,na49qm97,na49}. \\
But is it justified, based on these observations, to conclude that the observed spectra are the
result of a thermalised system and that the difference in (inverse) slopes for the various
particle species is a consequence of collective behaviour~?

\begin{figure}[htb]
\begin{center}
\resizebox{8.5cm}{!}{\includegraphics{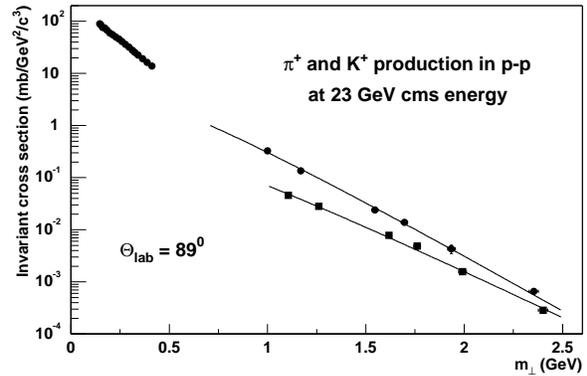}}
\end{center}
\caption{Charged pion ({\tiny \ding{108}}) and kaon ({\tiny \ding{110}}) yields observed in
         p-p collisions at $\sqrt{s}=23$~GeV at the CERN-ISR \cite{isr}.
         The solid lines indicate fits with the thermal model from Eq.~(\ref{eq:boltzmann2}).}
\label{fig:isr}
\end{figure}

To obtain an answer to the above questions, the charged pion and kaon yields at mid-rapidity as observed
in proton-proton collisions at the CERN-ISR \cite{isr} have been investigated as is
shown in Fig.~\ref{fig:isr}.     

From the CERN-ISR data it is seen that the pion and kaon spectra (similar results were
observed for $\pi^{-}$ and $K^{-}$ spectra) produced in proton-proton reactions at
similar collision energies as the CERN-SPS heavy-ion data are also compatible with our
calculated spectrum reflected by Eq.~(\ref{eq:boltzmann2}).
In addition, also here it is seen that particles with larger mass exhibit a larger inverse
slope in the $m_{\perp}$ spectrum.
The temperatures obtained here are 190~MeV and 223~MeV for the pions and kaons respectively,
which even quantitatively agree with the observations in the case of heavy-ion collisions
\cite{na49qm97}.
It should be noted that in \cite{na49qm97} a slightly different fit was used, but in the
$m_{\perp}$ range considered here this leads to only marginal differences in the extracted
temperatures.

Furthermore, even for high energy $e^{+}e^{-}$ annihilation reactions at the PETRA and LEP
accelerators \cite{lep} the particle spectra are found to match calculations based
on a thermal model.
Since it is very unlikely that in proton-proton and $e^{+}e^{-}$ collisions a thermalised
system will be created and collective expansion effects are present, the interpretation of the
observed $m_{\perp}$ spectra in terms of a thermalised and expanding hadronic source becomes
doubtful.\\
Further investigations to obtain a more detailed understanding of the underlying
reaction dynamics are therefore needed. Below, an attempt is made to explain the observed
distributions based on a common underlying production mechanism.

\section{How to understand the observations~?}
Due to the fact that the interactions described here are not calculable within QCD,
the interpretation of experimental results has to rely on phenomenological approaches.

\subsection{Statistical approach}
First attempts to interpret the experimental data of high energy nuclear collisions were performed
by trying to describe statistically the stage in which excitation of the incoming matter
takes place. However, in their original formulation \cite{stat1} these descriptions of
this so-called 'early stage' of the interaction failed to reproduce the experimental results. 

With the advent of more and more high quality data it turned out that after some 
generalisations had been applied to the original ideas, a statistical approach of the
early stage of the interaction \cite{marek} gives rather good agreement with the experimental results.\\
Furthermore, recently it has been demonstrated \cite{becattini,becattini2,jean} that also the hadronisation
process, which follows after the expansion and cooling of the early stage, can be described successfully
by a statistical model.  

Provided an extrapolation to a full phase-space coverage is performed, these statistical models
describe successfully the various hadron abundancies (i.e. the chemical composition)
of the interactions as observed in the various experiments.
However, they provide no description of the dynamical properties of the
created system as for instance reflected in the $m_{\perp}$ spectra of the reaction products.\\
These models also fail to describe the observed event topologies like the leading particle effect
and jet structures in $e^{+}e^{-}$ annihilations 
and in addition need the explicit assumption that the temperature of the system
remains constant during the complete hadronisation proces.
Furthermore, an ad-hoc strangeness suppression factor $(\gamma_{s})$ has to be introduced
to accommodate the observed yields of particles containing strangeness.

All this makes one wonder whether such statistical approaches are the right way to go.
Perhaps a description of the interaction process on a much more elementary level could
provide a proper description of all the observed characteristics.
 
\subsection{String fragmentation model}
Based on an intuitive interpretation of the Dual Resonance Model a string
model of hadrons has been developed \cite{string1}.
This model was seen to provide a simple explanation for confinement and also
leads to the fact that the particle spin is proportional to the square of the
particle mass, the so-called Regge behaviour.

Based on the success of the string model, such a string formalism was also 
used to provide a description of high energy collisions \cite{artru1,artru2} and
it was seen that such a phenomenological approach provided a quantitative
description of experimentally observed effects such as~:
\begin{itemize}
\item Leading particle effect and jet structures.
\item The existence of a rapidity plateau around mid-rapidity.
\item Limiting fragmentation, resulting in a logarithmic growth of the particle multiplicity
      as a function of the collision energy.
\item Relatively small values of the transverse momentum of the produced particles.
\end{itemize}  
Independently of the classical string phenomenology, a QCD inspired string fragmentation
model was developed \cite{lund} which also provided a quantitative description of
high energy interactions.  
The latter description has the advantage that it is well suited to be implemented 
in computer simulation programs, the so-called Lund Monte Carlo program.\\[5mm]
Both scenarios turn out to give similar results and in this paper the Lund
formalism will be used.  

\subsubsection{Transverse momentum spectra}
In section~5 of \cite{lund} it is demonstrated that fragmentation of a stretched-out
colour field provides a common mechanism to successfully describe particle production
for all of the following processes~:
\begin{itemize}
\item Quark fragmentation in $e^{+}e^{-}$ annihilations.
\item Meson fragmentation in hadronic collisions.
\item Proton fragmentation in hadronic collisions.
\item Proton fragmentation in deep inelastic scattering\\
      (diquark fragmentation).
\end{itemize}
In view of the similarities concerning the various $m_{\perp}$ spectra mentioned before,
it is therefore tempting to try to explain the observed $m_{\perp}$ distributions within
the framework of a string fragmentation model.

As outlined in \cite{lund}, hadron-hadron interactions can be described by representing
the initial hadrons as massless quarks connected by a string in "yo-yo mode" for a meson
and by a Y-shaped string for a baryon.
Due to the relative motion of the quarks in the initial hadron the colour field will not
be completely straight, but effectively in the overall cms of the collision a linear
description is obtained by projection on the direction $\vec{Q}+\vec{k}$, where $\vec{Q}$
represents the momentum transfer and $\vec{k}$ the primordial momentum of the struck (anti)quark.\\
Comparison of the string fragmentation model with various data indicates that the value of
$k_{\perp}$ (with respect to the direction $\vec{Q}+\vec{k}$) is of the order of about 400~MeV/c.
The remaining momentum $-\vec{k}$ of the system when the reaction takes place is distributed
among the particles in the "target" fragmentation region in which a particle with a large
energy-momentum fraction $z$ also takes a large fraction of this remaining momentum.

Limiting ourselves to soft processes and light-particle production at mid-rapidity
(i.e. low $z$ values), it follows that the effects of the primordial $k_{\perp}$
are marginal in our discussion concerning the observed $m_{\perp}$ spectra mentioned before.
In particular for initial baryons one quark will always be close to the junction which leads
to an essentially linear picture describing the diquark fragmentation along the incoming
beam direction. 

In the following the string fragmentation model will be used to derive the $m_{\perp}$
spectra for mesons produced at mid-rapidity in the case of hadron fragmentation.\\
Consider the case where, after string fragmentation, a quark of flavour $\alpha$ and an
anti-quark of flavour $\beta$ combine into a meson with transverse mass $m_{\perp}$ and
energy-momentum fraction $z$.
According to \cite{lund} the probability for this process to occur in the case of symmetric
string fragmentation is given by the so-called scaling function $f_{\alpha\beta}(z,m_{\perp})$
which is found to be
\begin{equation}
f_{\alpha\beta}(z,m_{\perp})=N_{\alpha\beta} (1-z)^{a} \, \frac{1}{z} \, e^{-bm_{\perp}^{2}/z} ~,
\label{eq:fab}
\end{equation}
where $a$ and $b$ are free parameters and $N_{\alpha\beta}$ represents a normalisation constant.\\
For low $z$ values the probability reflected in Eq.~(\ref{eq:fab}) can be written as
\begin{equation}
f_{\alpha\beta}(z,m_{\perp}) \sim \frac{1}{z} \, e^{-bm_{\perp}^{2}/z} ~,
\label{eq:fab2} 
\end{equation}
whereas for $z \rightarrow 1$ the expression of Eq.~(\ref{eq:fab}) basically becomes $(1-z)^{a}$.

Following the phenomenology outlined in detail in section~2.6 of \cite{lund} it is seen
that the parameters $a$ and $b$ are strongly correlated, since increasing $a$ or decreasing
$b$ will increase the mean multiplicity and affect the shape of the transverse momentum spectra.
In order to match the experimentally observed spectra and multiplicities only values along a curve
in the $(a,b)$ plane are allowed.\\
Good agreement with various data is obtained by $a=1$ and $b=0.4$~GeV$^{-2}$. 
Using these values of $a$ and $b$, it is seen that particle production at high $z$ values
is suppressed and that the distribution at low $z$ values peaks at $z=0.4 m_{\perp}^{2}$
(for low $m_{\perp}$ values) and vanishes for smaller $z$ values.

As discussed before, light-particle production at mid-rapidity due to soft processes
involves low $z$ values.
If we thus limit ourselves to this part of phase-space and make use of
$E=m_{\perp}\cosh(y)$, $p_{\parallel}=m_{\perp}\sinh(y)$ and
$z^{\pm}=(E \pm p_{\parallel})_{hadron}/(E \pm p_{\parallel})_{total}$,
the latter of which are the invariant lightcone variables describing the "forward" and
"backward" fragmentation in case of finite jet energies \cite{lund},
it is readily seen that in our case $z \sim m_{\perp}$.

This eventually leads to the following distribution function concerning light-particle
production in the central region~:
\begin{equation}
f(m_{\perp}) \sim \frac{1}{m_{\perp}} \, e^{\lambda m_{\perp}} ~,
\label{eq:fstring}
\end{equation}
where $\lambda$ comprises both the constant $b$ and the term $(E \pm p_{\parallel})_{total}$
corresponding to the forward or backward fragmentation.\\
From this probability the distribution of the particles emerging around mid-rapidity is found to be
\begin{equation}
 {\rm d}N(\vec{p},m) \sim
 e^{\lambda m_{\perp}}m_{\perp}\cosh(y)\,{\rm d}m_{\perp}{\rm d}y{\rm d}\varphi .
\label{eq:string}
\end{equation}
Here it is seen that also the phenomenological string fragmentation model results in
a transverse mass distribution of the particles produced in high energy collisions
which exhibits a shape compatible with a thermal source.\\
Obviously one could now raise the question of how Eq.~(\ref{eq:string}) would fit the
experimental spectra.

\begin{figure}[htb]
\begin{center}
\resizebox{8.5cm}{!}{\includegraphics{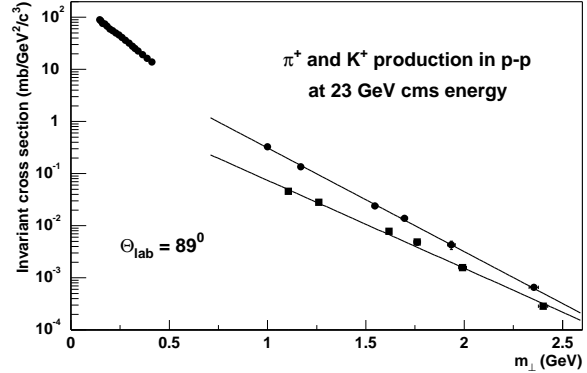}}
\end{center}
\caption{Charged pion ({\tiny \ding{108}}) and kaon ({\tiny \ding{110}}) yields observed in
         p-p collisions at $\sqrt{s}=23$~GeV at the CERN-ISR \cite{isr} fitted against
         the string fragmentation model from Eq.~(\ref{eq:string}).}
\label{fig:stringfit}
\end{figure}

Taking again the charged pion and kaon yields from the CERN-ISR proton-proton data as shown
in Fig.~\ref{fig:isr} and performing a fit of the data with Eq.~(\ref{eq:string}) shows that
the model describes the data rather well as can be seen from Fig.~\ref{fig:stringfit}.

Similar results are obtained for the $\pi^{-}$ and $K^{-}$ data.
The inverse slopes of the fit, which in a thermal model would represent the
temperature of the system, are found to be 218~MeV and 257~MeV for the pions and kaons, respectively.
As outlined in section 2.6 of \cite{lund}, also from the string fragmentation model it is seen that
the inverse slope becomes larger for heavier particles since they need a slightly higher
energy-momentum fraction $z$ at their creation.
Considering momentum values in the range of 0.5-1 GeV/c it is seen that kaons exhibit an increase
in $z$ of about 20\% compared to pions, resulting in a comparable increase in inverse slope of
the $m_{\perp}$ spectra.
This is perfectly in line with the observations in the case of the ISR data discussed above
and also in the case of heavy-ion collisions at the CERN-SPS \cite{na49qm97}.\\
Here it should be noted that the production of particles containing strangeness needs a
separate study concerning heavy-quark production in the string fragmentation model.
Since this is outside the scope of this paper we will not elaborate on this subject here,
but it is seen \cite{lund} that, apart from a relative suppression factor of about $\frac{1}{3}$
for strange-quark production, the fragmentation into kaons proceeds similarly to that into pions.\\
It is thus seen that the string model also automatically yields a suppression of heavy-quark production
which turns out to be in agreement with experimental observations and even matches the ad-hoc
strangeness suppression factor $\gamma_s=0.3$ which had to be introduced in the statistical
model \cite{becattini,becattini2,jean}. 

Comparison of the nucleus-nucleus data from Figs.~\ref{fig:na49} and \ref{fig:wa98} with the
string fragmentation model of Eq.~(\ref{eq:string}) is presented in the Figs.~\ref{fig:stringfit_na49}
and \ref{fig:stringfit_wa98} below.

\begin{figure}[htb]
\begin{center}
\resizebox{8.5cm}{!}{\includegraphics{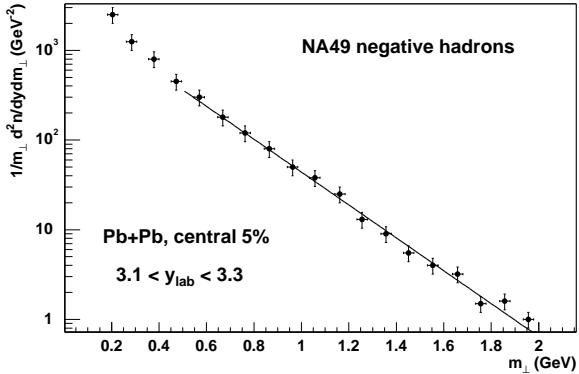}}
\end{center}
\caption{Charged hadron yields \cite{na49} as observed within the NA49 experiment.
         The solid line indicates a fit with the string fragmentation model from Eq.~(\ref{eq:string}).}
\label{fig:stringfit_na49}
\end{figure}

\begin{figure}[htb]
\begin{center}
\resizebox{8.5cm}{!}{\includegraphics{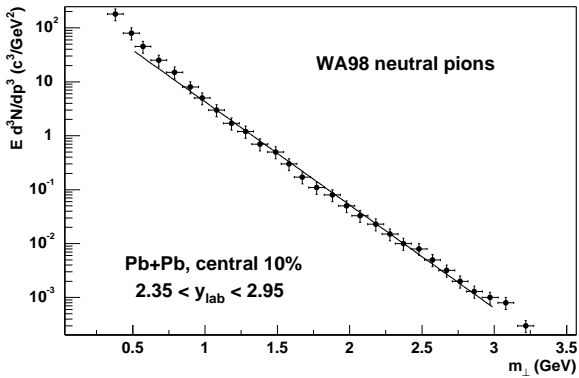}}
\end{center}
\caption{WA98 observed $\pi^{0}$ transverse mass spectrum \cite{wa98}.
         The solid line indicates a fit with the string fragmentation model from Eq.~(\ref{eq:string}).}
\label{fig:stringfit_wa98}
\end{figure}

Also in the case of the $m_{\perp}$ spectra of the hadrons produced in nucleus-nucleus
collisions the string fragmentation model is seen to describe the data rather well.
The inverse slopes from the NA49 and WA98 data are found to be 237~MeV and 227~MeV respectively.\\
In order to compare the various observed (inverse) slopes, the values obtained from the NA49
and WA98 data have to be corrected for the rapidity coverage.
Using the values of the specified rapidity intervals expressed in the overall cms of the collision 
and following the arguments given above, both inverse slopes amount to about 200~MeV after
the rapidity correction.\\
From this it is seen that indeed the string fragmentation model is able to describe all the
data discussed in this paper in a consistent way.

\section{Conclusions}
Various models describing particle production were confronted with data from
proton-proton and nucleus-nucleus data.\\
It was seen that both a thermal model and a phenomenological string fragmentation
model describe the observed transverse momentum spectra, where the string fragmentation
model automatically comprises the increase in inverse slope with particle mass as has been
observed in both the p-p and heavy-ion data discussed here.
To accommodate this in the thermal model, different freeze-out temperatures have to be
assumed for the various particle species or a collective expansion of the system has to be imposed,
in which the expansion velocity is the same for all particle species.

Concerning the production of various quark flavours, the string fragmentation model was seen
to yield automatically suppression of heavy quarks as compared to the light $u$ and $d$ flavours.
The strangeness suppression factor of about $\frac{1}{3}$ turned out to match the ad-hoc introduced
$\gamma_{s}$ from statistical models.\\
Finally it should be noted that the string model is the only one which accounts for the
observed event topologies like for instance the leading particle effect and jet structures.

From this it is concluded that a phenomenological string fragmentation model describes
all aspects of the data discussed here and consequently it is not justified to draw any
conclusions concerning a thermal nature or a collective expansion of the created system from
the observed transverse momentum spectra.     
     
\begin{ack}
The author would like to thank Michiel Botje, Adriaan Buijs and Marco van Leeuwen
for the inspiring discussions and valuable suggestions.\\
He would also like to acknowledge the ROOT team for providing such a powerful
analysis framework.
\end{ack}

\end{document}